\documentclass[aps,prb,twocolumn,
	           groupedaddress,superscriptaddress,
               amsfonts,amssymb,amsmath,floatfix,
	           citeautoscript]{revtex4-1}

\usepackage{graphicx}
\usepackage[centering,hmargin=20mm,tmargin=28mm,bmargin=25mm]{geometry}
\usepackage{newtxtext}
\usepackage[cmintegrals]{newtxmath}

\usepackage{siunitx}
\usepackage{xcolor}
\usepackage{hyperref}
\hypersetup{colorlinks,
	linkcolor={blue!75!black!80!yellow},
	citecolor={blue!75!black!80!yellow},
	urlcolor={blue!75!black!80!yellow}
}

\makeatletter
\renewcommand\@make@capt@title[2]{%
	\@ifx@empty\float@link{\@firstofone}{\expandafter\href\expandafter{\float@link}}%
	\sffamily{\textbf{#1}}\@caption@fignum@sep#2
}%

\makeatother

\usepackage{microtype} 
\usepackage{xspace}

\renewcommand{\Im}{\operatorname{Im}}
\renewcommand{\Re}{\operatorname{Re}}

\newcommand{\iu}{\mathrm{i}}
\newcommand{\e}{\mathrm{e}}
\newcommand{\dd}{\mathrm{d}}

\newcommand{\ie}{i.e.\@\xspace} 

\newcommand{\eg}{e.g.\@\xspace}

\newcommand{\HarvardSEAS}{John A. Paulson School of Engineering and Applied Sciences, Harvard University, Cambridge, MA, USA}
\newcommand{\MITPhy}{Department of Physics, Massachusetts Institute of Technology, Cambridge, MA, USA}

\usepackage{textcomp} 
\usepackage{xifthen}
\usepackage{etoolbox}
\newboolean{togglecomments}
\newboolean{togglechanges} 

\setboolean{togglecomments}{true}  
\setboolean{togglechanges}{false} 

\newcommand{\comment}[2]{%
    \ifbool{togglecomments}%
    {\textcolor{blue!70!black}{\small\textsf{%
    \textsuperscript{\textsc{\textsf{\MakeLowercase{#1}}}}%
    [#2]}}} 
    {}}     
\newcommand{\swap}[2]{\ifbool{togglechanges}
    {#2}  
    {\textcolor{red!70!black}{[#1]}\text{\textrightarrow{}}\textcolor{green!50!black}{[#2]}}}
\newcommand{\remove}[1]{\ifbool{togglechanges}
    {}    
    {\textcolor{red!70!black}{#1}}}
\newcommand{\inset}[1]{\ifbool{togglechanges}
    {{}#1}  
    {\textcolor{green!50!black}{{}#1}}}
\makeatletter
\newcommand{\citeremind}[1]{%
	\unskip%
    \textcolor{blue!75!black!80!yellow}{${}^\blacksquare$%
	\ifthenelse{\isempty{#1}}{}{\textsuperscript{\tiny\textsf{#1}}}%
	}\xspace}
\makeatother

\begin{document}

\title{Phonon polaritonics in two-dimensional materials}

\author{Nicholas Rivera}
\email{nrivera@seas.harvard.edu}
\affiliation{\HarvardSEAS}\affiliation{\MITPhy}
\author{Thomas Christensen}
\affiliation{\HarvardSEAS}\affiliation{\MITPhy}
\author{Prineha Narang}
\email{prineha@seas.harvard.edu}
\affiliation{\HarvardSEAS}

\date{\today}

\begin{abstract}
    Extreme confinement of electromagnetic energy by phonon polaritons holds the promise of strong and new forms of control over the dynamics of matter. 
    To bring such control to the atomic-scale limit, it is important to consider phonon polaritons in two-dimensional (2D) systems. Recent studies have pointed out that in 2D, splitting between longitudinal and transverse optical (LO and TO) phonons is absent at the $\Gamma$ point, even for polar materials.  
    Does this lack of LO--TO splitting imply the absence of a phonon polariton in polar monolayers? 
    Here, we derive a first-principles expression for the conductivity of a polar monolayer specified by the wavevector-dependent LO and TO phonon dispersions. In the long-wavelength (local) limit, we find a universal form for the conductivity in terms of the LO phonon frequency at the $\Gamma$ point,
    its lifetime, and the group velocity of the LO phonon. Our analysis reveals that the phonon polariton of 2D is simply the LO phonon of the 2D system. For the specific example of hexagonal boron nitride (hBN), we estimate the confinement and propagation losses of the LO phonons, finding that high confinement and reasonable propagation quality factors coincide in regions which may be difficult to detect with current near-field optical microscopy techniques. Finally, we study the interaction of external emitters with two-dimensional hBN nanostructures, finding extreme enhancement of spontaneous emission due to coupling with localized 2D phonon polaritons, and the possibility of multi-mode strong and ultra-strong coupling between an external emitter and hBN phonons. This may lead to the design of new hybrid states of electrons and phonons based on strong coupling.
   
\end{abstract}

\maketitle

Phonon polaritons, hybrid quasiparticles of photons and optical phonons supported in polar materials, hold promise for nanoscale control of electromagnetic fields at mid-infrared and terahertz frequencies. Qualitatively, phonon polaritons share many features with plasmon polaritons in conductors. Recently, it has been shown that phonon polaritons enable confinement of light to volumes  $\sim10^6$ times smaller than that of a diffraction-limited photon in free-space~\cite{caldwell2013low,xu2014mid,caldwell2014sub,dai2014tunable,tomadin2015accessing,yoxall2015direct,li2015hyperbolic,dai2015subdiffractional,dai2015graphene,caldwell2015low,li2016reversible,Basov:2016,basov2017towards,low2017polaritons,giles2017ultra,li2018infrared,ma2018plane}. Due to this remarkable confinement and their relatively high lifetimes---around picoseconds---phonon polaritons open new opportunities for vibrational spectroscopy~\cite{autore2018boron}, radiative heat transfer~\cite{hillenbrand2002phonon}, and control of dynamics in quantum emitters~\cite{kumar2015tunable,rivera2017making,kurman2018control,flick2018strong}. 

Thus far, extreme confinement of phonon polaritons has been achieved by the use of thin-films (or nanostructuring), which shrink the in- and out-of-plane wavelength of polaritons with decreasing feature size (such as the film thickness)~\cite{dai2014tunable,dubrovkin2018ultra}. A monolayer is the ultimate limit of this effect, making it critical to have a fundamental understanding of the optical response of 2D polar materials~\cite{thygesen2017calculating}. Concerning the optical response, the transition from three-dimensional (3D) to two-dimensional (2D) polar materials is nontrivial, however, since in a polar monolayer, the LO--TO splitting that gives rise to phonon polaritons in 3D is absent at the $\Gamma$ point~\cite{sanchez2002vibrational,mele2002electric,serrano2007vibrational,sohier2017breakdown}. This raises a fundamental question about the nature of  electromagnetic modes in polar monolayers.

We now develop a framework with quantitative, concrete examples that enables a first-principles understanding of phonon polaritons in 2D materials. 
We do so by deriving a universal form for the conductivity of a polar monolayer,  which depends solely on the LO and TO phonon frequencies---and their dispersion with momentum---in the 2D system. Using parameters from Ref.~\citenum{sohier2017breakdown} for the canonical 2D polar monolayer---hexagonal boron nitride (hBN)---we present the confinement and propagation losses of the 2D phonon polariton modes, identifying the frequency region where they should be most easily detected. Finally, we find that these modes enable extreme light-matter interaction between emitters and polar materials, showing that for atom-like emitters, their spontaneous decay can be enhanced by up to eight orders of magnitude through the emitter--LO phonon coupling. For an infrared emitter with a sufficiently high free-space radiative decay rate (${\gtrsim}\, \SI{1e6}{\per\s}$), we find that the associated linewidth of the emitter is comparable to the spacing between different phonon polaritonic resonances of an hBN nanostructure. This suggests the possibility of realizing the multi-mode strong coupling and ultra-strong coupling regimes of quantum electrodynamics in a 2D hBN platform. Our results for hBN are particularly relevant due its widespread use in 2D van der Waals heterostructures. In addition to providing functionality as a layer which improves the electrical and optical properties of other 2D materials, \eg graphene, our results suggest that in these heterostructures, hBN layers could provide a mid-infrared platform for nanophotonics and quantum optics. While we focus on hBN in this manuscript, the salient features of our findings apply to other polar monolayers as well.

\section{Optical response of optical phonons in two-dimensions}
In this section, we develop a theory of electromagnetic response due to optical phonons in 2D systems. The key response function of interest is the conductivity of the monolayer. To that end, we consider the response of the ions of the monolayer due to an electric potential $\phi$. For that case, the interaction Hamiltonian is
\begin{equation}
    H_{\mathrm{int}} = \int \dd^2x ~\rho \phi = -\int \dd^2x~ (\nabla\cdot\mathbf{P})\phi,
    \label{eq:coupling}
\end{equation} 
with $\rho$ the induced charge density and $\mathbf{P}$ the induced polarization density associated with the ionic motion. Note that boldfaced quantities refer to vectors or tensors as appropriate. Within linear response theory, the polarization density can be straightforwardly evaluated from the displacement  $\mathbf{u}_{\kappa}$ of every atom $\kappa$ within the unit cell. Specifically, to first order, the polarization density is 
\begin{equation}
\mathbf{P} - \mathbf{P}^0 = \sum_\kappa(\mathbf{u}_\kappa\cdot\boldsymbol{\nabla}_{\mathbf{u}_\kappa})\mathbf{P} \equiv \frac{1}{\Omega}\sum_\kappa\mathbf{Z}_{\kappa}\mathbf{u}_{\kappa},
\end{equation} 
where $\mathbf{Z}_\kappa \equiv \Omega\boldsymbol{\nabla}_{\mathbf{u}_\kappa}\mathbf{P}$ is the Born effective charge tensor of ion $\kappa$ and $\Omega$ is the unit cell area. $\mathbf{P}^0$ is the equilibrium polarization in the absence of displacements, which is zero here. With this relation between polarization and ionic displacements, the interaction Hamiltonian in Equation~\eqref{eq:coupling} couples the scalar potential and the ionic displacements.  We consider the response of the monolayer to a potential of the form $\phi(\mathbf{r}) = \phi(\mathbf{q},\omega)\e^{\iu\mathbf{q}\cdot\mathbf{r}-\iu\omega t}$, where $\mathbf{q}$ is a 2D wavevector in the plane of the monolayer. Such a potential corresponds to a longitudinal electric field  $\mathbf{E}(\mathbf{r}) = \iu\mathbf{q}\phi(\mathbf{r})$. 

In what follows, we assume the validity of the random-phase approximation (RPA) in calculating Coulombic interactions between ions in the polar lattice. Within the RPA, these Coulombic interactions are accounted for by taking the induced polarization $\mathbf{P}(\mathbf{q},\omega)$ to be proportional \emph{total} electric field, $\mathbf{E}_{\mathrm{tot}}(\mathbf{q},\omega)$, defined to be the sum of the externally applied electric field and the electric field created by the induced polarization. The polarization and total field are connected by the polarization-polarization response (tensor) function  $\boldsymbol{\Pi}(\mathbf{q},\omega)$ via
\begin{equation}
\mathbf{P}(\mathbf{q},\omega)  = \epsilon_0\boldsymbol{\Pi}(\mathbf{q},\omega)\mathbf{E}_{\mathrm{tot}}(\mathbf{q},\omega),
\end{equation}
The polarization-polarization response function is related to the conductivity via the relation $\boldsymbol{\sigma}(\mathbf{q},\omega) = -\iu\omega\epsilon_0\boldsymbol{\Pi}(\mathbf{q},\omega)$. From the Kubo formula, it follows that the conductivity is
\begin{equation}\label{eq:2dsusceptibility}
    \boldsymbol{\sigma}(\mathbf{q},\omega) =  \frac{-\iu \omega}{ \Omega\mathcal{Z}}\sum\limits_{m,n}\frac{\mathbf{P}_{mn}(\mathbf{q})\otimes\mathbf{P}_{nm}(\mathbf{q})}{\hbar\omega + E_{nm}+\iu 0^+}\Big(\e^{-\beta E_m}-\e^{-\beta E_n} \Big),
\end{equation}
where $m,n$ are eigenstates in the phononic Fock space of the monolayer, $\mathbf{P}_{mn}(\mathbf{q}) \equiv \sum_\kappa \mathbf{Z}_\kappa\langle m|\mathbf{u}_\kappa(\mathbf{q})|n\rangle$ are matrix elements of the polarization associated with phonon modes (where $\mathbf{u}_\kappa(\mathbf{q})$ is the Fourier transform of the phonon displacement operator), $E_{m}$ ($E_n$) is the energy of state $m$ ($n$), $\beta \equiv 1/k_{\text{\textsc{B}}}T$ is the inverse temperature, and $\mathcal{Z}$ is the grand partition function. We now evaluate the contribution of optical phonons to the polarization-polarization response in the low-temperature limit $T \ll \hbar\omega_{\mathrm{ph}}/k_{\mathrm{B}}$ with $\omega_{\mathrm{ph}}$ a characteristic optical phonon frequency.  Considering the long-wavelength (small wavevector) limit, and taking a material with long-wavelength isotropy, such as hBN, we only have to consider the $qq$-component in the response tensor, where $qq$ denotes a pair of directions parallel to the wavevector. Denoting $\sigma_{qq}$ as simply $\sigma$, we find that the conductivity is given by~\cite{rivera2018ab}: 
\begin{equation}
    \sigma(\mathbf{q},\omega) = \frac{-\iu\omega}{\hbar\Omega} \frac{2\omega_{\mathbf{q},\mathrm{L}}}{\omega^2_{\mathbf{q},\mathrm{L}}-\omega^2-\iu\omega\tau^{-1}}|\hat{\mathbf{q}}\cdot\langle 1_{\mathbf{q},\mathrm{L}}|\mathbf{P}(\mathbf{q})|0_{\mathbf{q},L}\rangle|^2,
    \label{eq:polpolresponse}
\end{equation}
where L-subscripts denote longitudinal polarization, $|0_{\mathbf{q},\mathrm{L}}\rangle$ ($|1_{\mathbf{q},\mathrm{L}}\rangle$) denotes a state with no (one) longitudinal phonon of wavevector $\mathbf{q}$, and $\hat{\mathbf{q}}$ denotes a unit vector in the direction of $\mathbf{q}$. We have also phenomenologically included the phonon dissipation rate $\tau^{-1}$, consistently with  a relaxation-time prescription.  The frequency $\omega_{\mathbf{q},\mathrm{L}}$ in the denominator, as in the case of bulk phonons, is the frequency of the longitudinal phonon of wavevector $\mathbf{q}$ prior to considering LO--TO splitting~\cite{BornHuang:1954} (and near the $\Gamma$ point is approximately equal to the TO phonon frequency). This is consistent with the fact that LO--TO splitting is a collective effect arising from Coulomb interactions and the fact that the equation above represents a single-particle susceptibility. Coulomb interactions are accounted for in the random phase approximation, and to include them in the single-particle response amounts to an uncontrolled double-counting. 

Next, we express the polarization matrix element in Equation~\eqref{eq:polpolresponse} in terms of the Born effective charges of the monolayer and the phonon displacement eigenvectors. Considering the longitudinal phonon contribution to the second-quantized ionic displacement, as in Ref.~\citenum{srivastava1990physics}, we find that the conductivity within the RPA is given by
\begin{equation}
    \sigma(\mathbf{q},\omega) = -\frac{\iu\omega }{\Omega}\frac{\Big|\hat{\mathbf{q}}\cdot\sum\limits_{\kappa}\mathbf{Z}_{\kappa}\boldsymbol{\eta}_{\kappa} \Big|^2}{\omega^2_{\mathbf{q},\mathrm{L}}-\omega^2-\iu\omega\tau^{-1}}.
    \label{eq:conductivity_simplified}
\end{equation} 
We have defined scaled eigendisplacements $\boldsymbol{\eta}_{\kappa\mathbf{q}}\equiv \frac{\hat{\mathbf{e}}_{\kappa\mathbf{q},\mathrm{L}}}{\sqrt{M_{\kappa}}}$, where $\hat{\mathbf{e}}_{\kappa\mathbf{q},\mathrm{L}}$ is the unit-normalized polarization vector of atom $\kappa$ in the unit cell oscillating according to a longitudinal phonon of wavevector $\mathbf{q}$ and $M_{\kappa}$ is the mass of atom $\kappa$. 

While the conductivity is the main electromagnetic quantity of interest for electrodynamics applications, we briefly state the form of the (2D) permittivity, as its zeros immediately yield the longitudinal modes of the system, which are the LO phonons. The permittivity within the RPA, denoted $\epsilon_{\textrm{RPA}}$ is related to the polarization-polarization response function via~\cite{jablan2009plasmonics} $\epsilon_{\textrm{RPA}} = \epsilon_{\mathrm{env}} + \frac{1}{2}q\Pi(q,\omega)$ and the conductivity via $\epsilon_{\textrm{RPA}} = \epsilon_{\mathrm{env}} + \frac{\iu q\sigma(q,\omega)}{2\omega\epsilon_0}$. Here, $\epsilon_{\mathrm{env}}$ is the average permittivity of the bulk above and below the monolayer, and is added to take into account the polarization arising from these bulk materials. Note that we have neglected any intrinsic high-frequency screening in the monolayer itself, which is only relevant for wavevectors comparable to the inverse layer spacing between monolayers. When considering non-local corrections to the conductivity at these large wavevectors, these must be taken into account~\cite{sohier2016two,thygesen2017calculating,sohier2017breakdown}. Based on Equation~\eqref{eq:conductivity_simplified}, the zeros $\omega_{\mathbf{q}}$ of the RPA dielectric function satisfy:
\begin{equation}
    \omega^2_{\mathbf{q}} - \omega^2_{\mathrm{TO}} = \frac{V(q)}{e^2}\frac{1}{\Omega}q^2\Big|\hat{\mathbf{q}}\cdot\sum\limits_{\kappa}\mathbf{Z}_{\kappa}\boldsymbol{\eta}_{\kappa}\Big|^2,
    \label{eq:zerosofresponse}
\end{equation}
  where $V(q)$ is the Coulomb interaction in Fourier space, which in two dimensions, is given by $V(q) = \frac{e^2}{2\epsilon_0\epsilon_{\mathrm{env}}q}$. Given that the zeros of the dielectric function are associated with longitudinal modes, one expects that $\omega_{\mathbf{q}}$ is in fact the frequency of the LO phonon mode. This is consistent with the result of Ref.~\citenum{sohier2017breakdown}, in which \relax{{}\ }\citeauthor{sohier2017breakdown} show that in 2D polar materials, the extra restoring forces on LO phonons relative to TO phonons, due to the Coulomb interaction, lead to a wavevector-dependent LO--TO splitting and zero LO--TO splitting at the $\Gamma$ point of the Brillouin zone. 
  
Given these results, we now re-express the conductivity explicitly in terms of the 2D phonon dispersion, and derive a universal form for the conductivity in the local ($q\rightarrow 0$) limit specified in terms of three parameters: the LO phonon frequency at the $\Gamma$ point (i.e., $\omega_{\mathrm{TO}}$), the group velocity of the LO phonon at the $\Gamma$ point, and the damping rate. From Equation~\eqref{eq:zerosofresponse}, we can immediately write the conductivity as
  \begin{equation}
  \sigma(\mathbf{q},\omega) = \frac{-2\iu\epsilon_0\epsilon_{\mathrm{env}}\omega}{q}\frac{\omega^2_{\mathbf{q},\mathrm{LO}}-\omega^2_{\mathrm{TO}}}{\omega^2_{\mathrm{TO}}-\omega^2-\iu\omega\tau^{-1}}.
  \end{equation}
 In this expression, $\omega_\mathrm{L}$, the LO phonon frequency prior to LO-TO splitting, has been re-named as $\omega_{\mathrm{TO}}$ , the transverse optical phonon frequency, because in the absence of LO--TO splitting, they are degenerate. The RPA zeros $\omega_{\mathbf{q}}$ have also been renamed as $\omega_{\mathbf{q},\mathrm{LO}}$. This is done in order to make the form of the final results more closely resemble their 3D counterparts, in which the dielectric function is expressed in terms of the TO frequency (see for example Eq. (\ref{eq:3deps})). 
 
 For small $q$,  the Born charges are (to lowest-order) constant, and so the LO phonon dispersion takes the form $\omega_{\mathbf{q},\mathrm{LO}}=\sqrt{\omega^2_{\mathrm{TO}}+2v_{\mathrm{g}}\omega_{\mathrm{TO}}q} \simeq \omega_{\mathrm{TO}} + v_{\mathrm{g}}q$, where $v_{\mathrm{g}}$, the LO phonon group velocity, is defined from microscopic parameters through the relation
 \begin{equation}
    v_{\mathrm{g}} = \frac{\Big|\hat{\mathbf{q}}\cdot \sum_\kappa \mathbf{Z}_\kappa\boldsymbol{\eta}_\kappa\Big|^2 }{4\epsilon_0 \epsilon_{\mathrm{env}}\omega_{\mathrm{TO}}\Omega}. 
\end{equation}
Thus, in the long wavelength limit, we have the following universal parameterization of the conductivity of a polar monolayer:
\begin{equation}
    \sigma(\omega) =  \frac{-4\iu\epsilon_0\epsilon_{\mathrm{env}}\omega\omega_{\mathrm{TO}}v_{\mathrm{g}}}{\omega^2_{\mathrm{TO}}-\omega^2-\iu\omega\tau^{-1}}.
    \label{eq:localsigma}
\end{equation}

We note that despite its appearance, $\sigma(\omega)$ does not depend on $\epsilon_{\mathrm{env}}$, as $v_{\mathrm{g}}$ has an opposite dependence on $\epsilon_{\mathrm{env}}$. From this relation, it follows that given the properties of the 2D phonons (from experiments or from \emph{ab initio} calculations), one can immediately specify the conductivity. Alternatively, from optical measurements (including far-field measurements) which allow one to extract the conductivity, it becomes possible to extract the group velocity of 2D LO phonons and thus the small-wavevector dispersion of those phonons. 
  
Before moving on to analyze the electrodynamics of 2D phonon polaritons, we make two comments on lack of LO--TO splitting in 2D polar materials. The first is that this situation is in stark contrast to the situation of polar materials in 3D, which have a finite LO--TO splitting at the $\Gamma$ point. In the absence of such LO--TO splitting in 3D, there would be no frequency compatible with the existence of a phonon  polariton. On the contrary, we will show that in 2D, despite the absence of LO--TO splitting at the $\Gamma$ point, there persists a strongly confined evanescent electromagnetic mode with a high local density of states which in all respects is similar to a phonon polariton of a thin film, but is in fact the 2D LO phonon of the polar monolayer (thus the phrases ``phonon polariton'' and ``2D LO phonon'' may be used somewhat interchangeably as is the case in plasmonics where the terms ``plasmon polariton'' and ``2D plasmon'' are often used interchangeably).

The second comment is that much of what has been discussed here has a strong analogy with the theory of optical response in electron gases in 2D, and particularly the relation between plasmons in 2D and 3D. To elaborate on this analogy, we take Equation~\eqref{eq:zerosofresponse} in the case of a two-atom unit cell (such as hBN), and note that the term in the sum over Born charges can be written as $\Big|\hat{\mathbf{q}}\cdot \sum_\kappa \mathbf{Z}_\kappa\boldsymbol{\eta}_\kappa\Big|^2 \equiv \frac{Q_*^2}{M_*}$, with $Q_*$ being an effective charge and $M_*$ being an effective mass. Then, the LO--TO splitting can be written as $\omega^2_{\mathbf{q}} - \omega^2_{\mathbf{q},\mathrm{TO}} = \frac{Q_*^2}{2\epsilon_0\epsilon_{\mathrm{env}} M_*}q$. Now we note that the RHS is exactly the squared-frequency $\omega^2_{\mathbf{q}\mathrm{p}}$ for a plasma oscillation in a 2D gas of charged particles with charge $Q_*$ and mass $M_*$. To connect to LO-TO splitting in phonons, this squared frequency $\omega^2_{\mathbf{q}\mathrm{p}}$ can be thought of as the ``LP--TP'' splitting between longitudinal and transverse plasma oscillations. Of course, there are no transverse plasma oscillations due to the structure-less nature of the electron, and so ``$\omega_{\mathrm{TP}}$''  should be considered equal to zero. In the three-dimensional plasmon case, ``$\omega_{\mathrm{TP}} = 0$'', but the difference between the squared longitudinal and transverse plasma oscillation frequencies at zero-wavevector is non-zero and given by $\omega_{\mathrm{p}}^2$. In other words, the plasma frequency in electron gases is analogous to the LO--TO splitting in polar materials. The change in the dispersion of 3D versus 2D plasmons, like 3D versus 2D LO phonons, arises from the change in dimensionality of the Coulomb interaction from 3D to 2D.  This analogy between phononic and plasmonic behavior as a function of dimension is illustrated schematically in Figures~\ref{fig:1}(b,c) in order to help unify the understanding of 2D plasmonics and phononics. These considerations should also extend to the one-dimensional case.
  
\section{Electrodynamics of optical phonons in two-dimensions}
\begin{figure}[t]
    \includegraphics[width=0.95\linewidth]{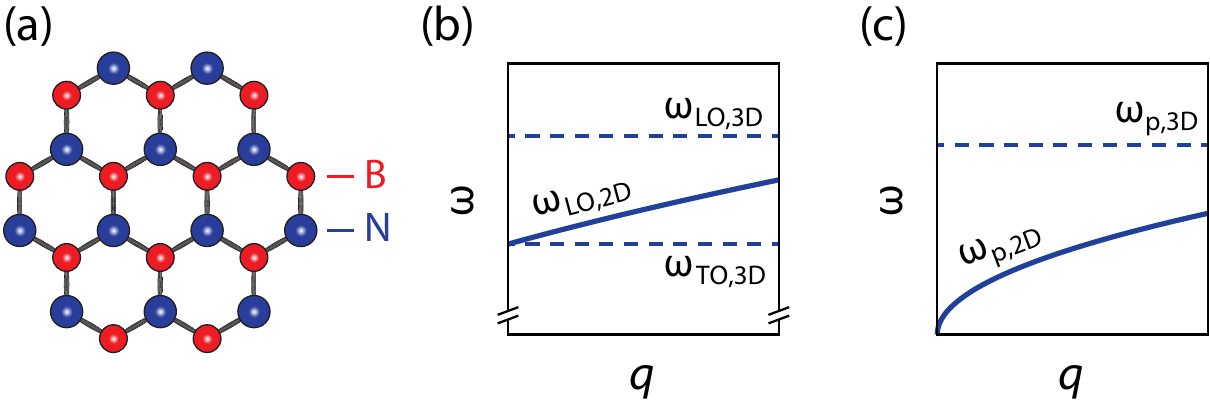}
    \caption{%
        \textbf{LO phonons as the basic electromagnetic waves of a polar monolayer.} (a) Schematic structure of a polar monolayer such as hexagonal boron nitride. (b) Properties of LO and TO phonons in 3D and 2D. In 3D, there is a finite LO--TO splitting at zero wavevector, while in 2D there is none. Despite this, the 2D LO phonon plays the role of the phonon polariton in 3D and thin films. (c) Analogous physics appears in electron gases in 3D and 2D, where the 3D plasma frequency is similar to the 3D LO--TO splitting. In 2D, the plasma frequency at zero wavevector is zero, but the electromagnetic physics is determined by the dispersion of 2D plasmons, which replace the plasmon polariton of bulk and thin films.
        \label{fig:1}
        }
\end{figure}

To relate the conductivity function  to the electromagnetic modes supported by a polar monolayer, we solve Maxwell's equations for an evanescent electromagnetic mode supported by a surface with conductivity $\sigma$. We consider the monolayer to be sandwiched by a superstrate of permittivity $\epsilon_+$ and a substrate of permittivity $\epsilon_-$. To strip the analysis to its bare essentials, we consider optical phonon response with in-plane isotropy in the long-wavelength limit arising from in-plane LO oscillations. A relevant example of a system where these conditions are satisfied is in a hexagonal boron nitride monolayer (see Figure~\ref{fig:1}(a) for schematic atomic structure). In a monolayer geometry with translation invariance and in-plane isotropy, the solutions of Maxwell's equations can be decomposed into transverse magnetic (TM) and transverse electric (TE) parts, where the magnetic or electric field respectively is transverse to the in-plane wavevector of the mode. In practice, it is the TM mode which is associated with highly confined electromagnetic waves.   We consider without loss of generality a TM mode with wavenumber $q$ along the $x$-direction in the monolayer and magnetic field $H(z)\e^{\iu qx-\iu\omega t}$ along the $y$-direction of the monolayer. The direction transverse to the monolayer is denoted as $z$.  With these definitions in place, the Maxwell equation satisfied by the magnetic field is 
\begin{equation}\label{eq:2dmaxwell}
    \bigg(-\frac{\dd^2}{\dd{}z^2}+q^2-\epsilon_{\pm}\frac{\omega^2}{c^2} \bigg)H(z) = 0,
\end{equation}
where $\epsilon_+$ applies for $z>0$ and $\epsilon_-$ applies for $z<0$.
We consider a solution of the form $H(z) = h_\pm\e^{\mp\kappa_\pm z}$ with $\kappa_\pm = \sqrt{q^2-\epsilon_{\pm} \frac{\omega^2}{c^2}}$ with $\pm$ corresponding to $\pm z > 0$ respectively.  The boundary condition on the magnetic field is $h_+ - h_- = -K_x = -\sigma E_x$ where $\mathbf{K}$ is the surface current density, and $\mathbf{E} = -\frac{1}{i\omega\epsilon}\nabla\times\mathbf{H}$ is the electric field. This condition enforces $h_+-h_- = \frac{\sigma}{i\omega\epsilon_+}\kappa_+h_+$. 
Continuity of the electric field in the $x$ direction enforces ${\epsilon_-}/{\epsilon_+} = -{\kappa_-h_-}/{\kappa_+h_+}$. 
Combining the two conditions, we obtain the usual dispersion equation for the TM mode of a polarizable 2D monolayer, namely $\epsilon_+/\kappa_+ + \epsilon_-/\kappa_- = \frac{\sigma}{i\omega}$. Given the deeply subwavelength nature of 2D phonon polaritons, \ie since $q\gg\omega/c$ such that $\kappa_{\pm}\simeq q$, the dispersion equation can be reduced to its quasistatic limit without consequential loss of accuracy:
\begin{equation}
 q=\frac{2i\omega\epsilon_0\epsilon_{\mathrm{env}}}{\sigma},
    \label{eq:dispeq}
\end{equation}
with $\epsilon_{\mathrm{env}}\equiv (\epsilon_++\epsilon_-)/2$. This condition, as can be seen from the relation between $\Pi$ and $\sigma$, is equivalent to $\epsilon_{\mathrm{RPA}}=0$. Thus, the condition for phonon polaritons coincides precisely with the condition for longitudinal optical phonons.  One of the main results of our manuscript is that despite the lack of LO--TO splitting at the $\Gamma$ point, there nevertheless exists a strongly confined evanescent mode in many respects similar to the phonon polaritons of thin films. We now analyze the dispersion relation of phonon polaritons in a specific material, hexagonal boron nitride, in more detail, showing the possibility of highly confined electromagnetic modes with a large local density of states. 

~

\begin{figure*}[t]
    \includegraphics[width=.7\linewidth]{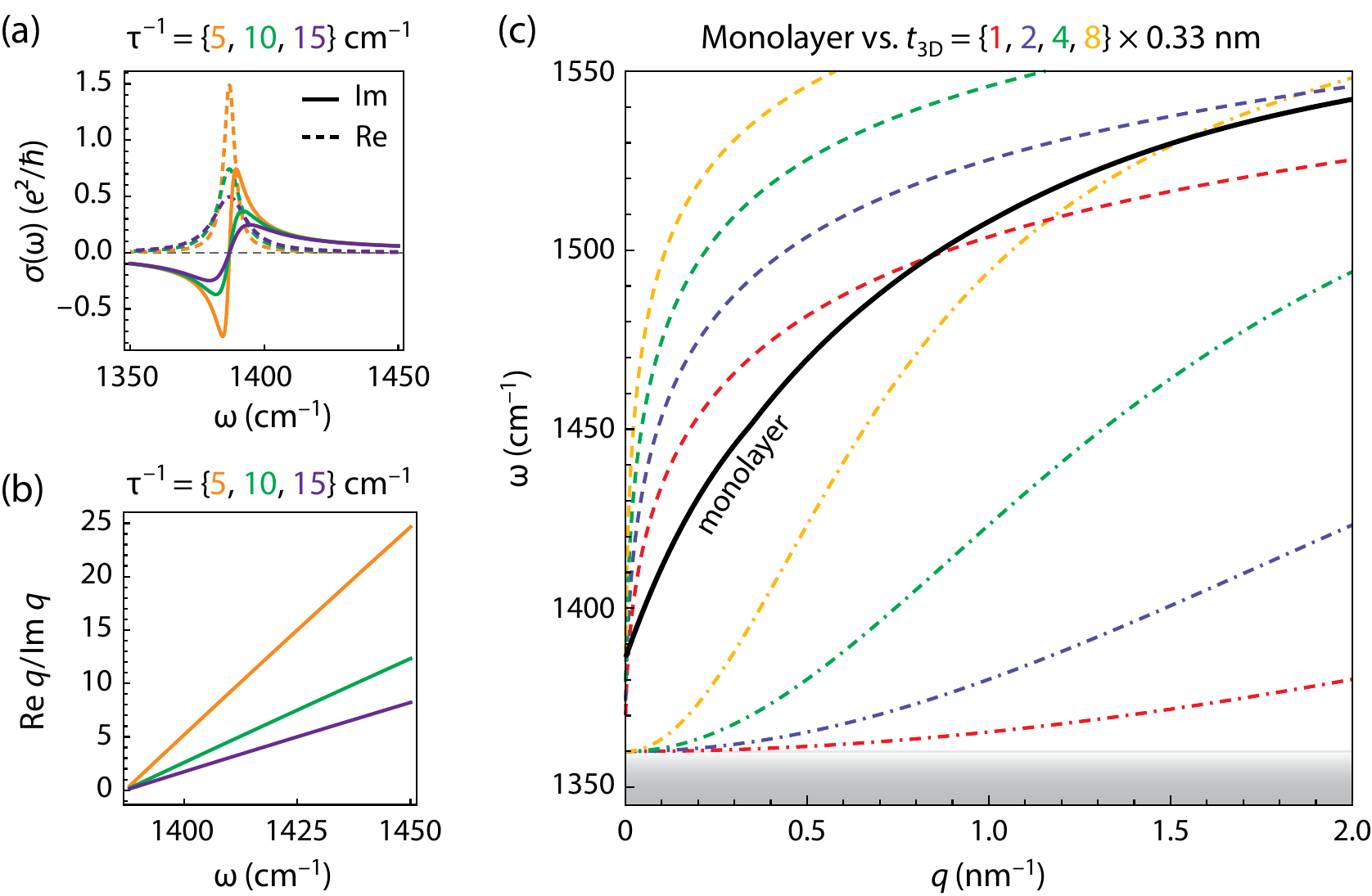}
    \caption{%
        \textbf{Properties of phonon polaritons in a monolayer of hBN compared to bulk.} (a)~Real and imaginary parts of the conductivity of 2D hBN for different values of the loss-rate. (b)~Propagation quality factor, which measures the number of wavelengths of propagation of the 2D phonon polariton. (c)~Dispersion relation of phonon polaritons in the monolayer (black) and thin films whose thicknesses are taken to be 1, 2, 4, and 8 interlayer spacings in hBN. For the thin-film, the fundamental mode (dashed) and the first higher-order mode (dash-dotted) are plotted. 
        \label{fig:2}
        }
\end{figure*}

In Figure ~\ref{fig:2}(a), we present the conductivity of 2D hBN of Equation~\ref{eq:localsigma}, using parameters from Ref.~\citenum{sohier2017breakdown} calculated from density functional theory within the local density approximation. In this plot $\epsilon_{\mathrm{env}}=1,$ $\omega_{\mathrm{TO}} = $ \SI{1387}{\per\cm} and $v_{\mathrm{g}} = 1.2\times 10^{-4}c$, with $c$ the speed of light in vacuum.

From the conductivity, the dispersion relation of phonon polaritons on an infinite sheet is given by $q={2\iu\omega\epsilon_0\epsilon_{\mathrm{env}}}/{\sigma(\omega)}$. The dispersion, assuming $\tau = \infty$ and $\epsilon_{\mathrm{env}}=1$, is shown in Figure~\ref{fig:2}(c) (black line). A key figure of merit for applications involving the propagation of phonon polaritons, is the propagation quality factor, defined by ${\Re q}/{\Im q} = {\Im\sigma(\omega)}/{\Re\sigma(\omega)}$, which is shown in Figure~\ref{fig:2}(b).

For monolayer hBN, the wavevector grows very rapidly with frequency, due to the extremely low group velocity of 2D LO phonons, which is a remarkable four orders of magnitude slower than the speed of light. In particular, at frequencies of \SI{1450}{\per\cm}, the phonon polariton has a wavelength of about \SI{15}{\nm}, significantly shorter than any phonon polariton measured so far, and, similarly shorter than any plasmonic wavelength, even in graphene. In fact, this short a wavelength well-below that of any polariton in current scattering near-field microscopy (SNOM) measurements. The 2D phonon polariton could in principle be measured by SNOM closer to the TO frequency, where confinement is smaller; unfortunately, as shown in Figure~\ref{fig:2}(b), near the TO frequency, dissipation is far higher (and corresponding propagation quality factors $\Re q/\Im q$ far lower) due to large $\Re\sigma$ (or, equivalently, large $\Im\epsilon_{\mathrm{RPA}}$).

These considerations imply that access to the lower-loss and higher-confined portions of the dispersion relation of phonon polaritons, in the absence of a sharper tip, requires a near-field probe such as a free electron probe, as used in electron energy loss spectroscopy (EELS), where slow electrons can be used to probe plasmon wavelengths of just a few nanometers in monolayer metals, as well as the nonlocal bulk plasmon dispersion in metals~\cite{nagao2001dispersion,de2010optical,diaconescu2007low}.
EELS has been recently employed to measure phonon polaritons in ultrathin films of hBN~\cite{govyadinov2017probing}. Another interesting class of near-field probes, with relevance to fundamental physics and quantum optics applications, is a quantum emitter such as an atom, molecule, or artificial atom such as a quantum dot, quantum well, or vacancy center. Recently, it was demonstrated using nanostructures of bulk hBN that the interaction of vibrational emitters with phonon polaritons is on the border of the strong coupling regime~\cite{autore2018boron}.

In the rest of this section, we discuss the relation between the dispersion of an hBN monolayer versus the atomically-thick limit of a thin film of a material with hBN's bulk dielectric function. To aid this discussion, in Figure~\ref{fig:2}(c), we show the dispersion relation of thin films of bulk hBN with film thicknesses of 1, 2, 4, and 8 times the interlayer spacing of bulk hBN, which is roughly \SI{0.33}{\nm}. For these plots, we take hBN to be cleaved such that the optical axis is perpendicular to the plane of the film. The components of the bulk permittivity perpendicular and parallel to the c-axis ($\epsilon_{\perp}$ and $\epsilon_{\parallel}$, respectively; indexed by $\alpha\in\{\perp,\parallel\}$ below) are then given by
\begin{equation}
	\epsilon_{\alpha}(\omega) = \epsilon_{\infty,\alpha}\Bigg(1 + \frac{\omega^2_{\mathrm{LO},\alpha}-\omega^2_{\mathrm{TO},\alpha}}{\omega^2_{\mathrm{TO},\alpha}-\omega^2} \Bigg),
	\label{eq:3deps}
\end{equation}
with $\epsilon_{\infty,\parallel} = \num{2.95}$, $\omega_{\mathrm{TO},\parallel} = \SI{760}{\per\cm}$, and $\omega_{\mathrm{LO},\parallel} = \SI{830}{\per\cm}$; and $\epsilon_{\infty,\perp} = \num{4.87}$, $\omega_{\mathrm{TO},\perp} = \SI{1360}{\per\cm}$, and $\omega_{\mathrm{LO},\perp} = \SI{1614}{\per\cm}$\cite{caldwell2014sub,dai2014tunable}.
Losses are ignored in this discussion altogether. 
In the range between $\omega_{\mathrm{TO}, \alpha}$ and $\omega_{\mathrm{LO}, \alpha}$, the corresponding component of the permittivity is negative, while the other component is positive. This hyperbolicity leads to a dispersion for hBN thin films that have multiple branches at a given frequency, as can be seen in Figure~\ref{fig:2}(c).  This trend persists even when the thickness of the bulk is taken down to a single layer, albeit being pushed to high wavevectors. This is in contrast with the true monolayer, in which there is only one LO phonon mode. Given that the phonon polariton of 2D is the LO phonon, there can be only one branch of the dispersion. Another anomaly between 2D and 3D is that the dispersion relation of the 2D LO phonon is non-monotonic, and the frequency starts to decrease for sufficiently large wavevector. This can never happen by extrapolating to a monolayer a dielectric function which has finite LO--TO splitting. This turn-back point corresponds to a frequency of around \SI{1500}{\per\cm} and a phonon wavelength of less than \SI{1}{\nm}. Thus, this deviation is challenging to probe in typical optical experiments. However it means that for the monolayer, the range of accessible phonon polaritonic frequencies for optical applications is reduced compared to thick films. Finally, while we have focused on anomalies between the atomically-thin limit of bulk and a true monolayer, a comparison of the monolayer with the fundamental phonon polariton mode of the one-atom-thick thin film suggests that we have a reasonable qualitative understanding of the monolayer dispersion from this limit.

\begin{figure}[t]
    \includegraphics[width=.925\linewidth]{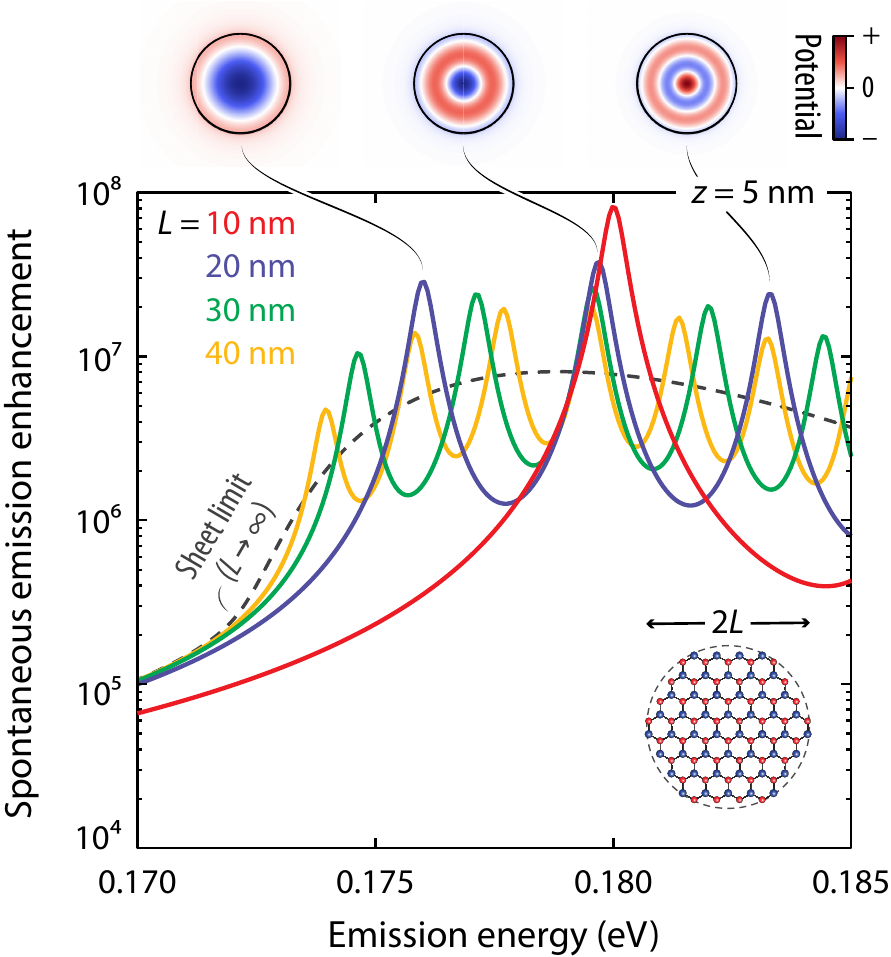}
    \caption{%
        \textbf{Extreme spontaneous emission enhancement due to 2D phonon polaritons in nanostructured geometries.} Plotted is the enhancement of the spontaneous emission rate for an emitter  $z=\SI{5}{\nm}$ above the disk's center and  oriented normal to its plane ($\hat{\mathbf{n}} = \hat{\mathbf{z}}$). For a disk with a diameter of \SI{20}{\nm}, and assuming a relaxation rate $\tau^{-1} = \SI{0.5}{\meV}$, the rate of emission enhancement can be enhanced \num{100} million-fold. For an emitter with a free-space decay rate of $\SI{1e6}{\per\cm}$ at \SI{7}{\micro\m}, the emitter would experience a decay rate comparable to the frequency of the disk mode, leading to ultra-strong coupling of an external emitter with 2D phonon polaritons. For reference, we compare the spontaneous emission enhancement in a nanodisk geometry to that of a disk of infinite radius (i.e., a flat sheet), showing clear enhancement relative to the flat sheet due to concentration of local density of states around phonon polariton resonances. Note that $\tau^{-1} = \SI{0.5}{\meV}$ is of the order of the loss rate in bulk hBN. Also shown in the figure are maps of the electric potential on the surface of the disk for modes corresponding to selected phonon polariton resonances in the plot. 
        \label{fig:3}
        }
\end{figure}

\section{Strong light-matter interactions enabled by 2D optical phonons}

The extreme confinement of electromagnetic fields offered by the 2D phonon polariton presents an opportunity for quantum optical applications in which one seeks to couple an external emitter such as an atom, molecule, defect, or artificial atomic system to electromagnetic fields. Applications of these couplings are ultra-bright single- or two-photon sources, realizing the strong-coupling regime and the associated phenomenology of Rabi oscillations and polaritons, or resolving spectroscopically ``forbidden'' transitions~\cite{koppens2011graphene,autore2018boron,rivera2016shrinking,rivera2017making,kurman2018control} to achieve near-field spectroscopies with momentum and angular momentum resolution not accessible in far-field spectroscopies.

In Figure~\ref{fig:3}, we consider the coupling of a dipole emitter to localized phonon polaritons of nanostructured monolayer hBN. For simplicity, we consider hBN nanostructured as a disk, which leads to the formation of sharp resonances quantized along the azimuthal and radial directions. The disk is taken to have a radii varying from \SI{10}{\nm} to \SI{40}{nm}
and a loss rate $\tau^{-1} = \SI{0.5}{\meV}$, which is of the order of the loss rate in bulk hBN~\cite{caldwell2014sub, dai2014tunable}. We also show (dashed line) the results for a disk of infinite radius, i.e., a flat sheet of monolayer hBN. We parameterize the coupling between the dipole and phonon polaritons through the rate of spontaneous emission $\Gamma$ of phonon polaritons by the dipole, normalized to the rate of spontaneous emission in free space $\Gamma_0$. It is related to the dyadic Green function $\mathbf{G}$ of the Maxwell equations for the nanostructure via the expression~\cite{novotny2012principles}:
\begin{equation}
	\frac{\Gamma}{\Gamma_0} 
	=  \frac{6\pi c}{\omega}\hat{\mathbf{n}}\cdot\Im\mathbf{G}(\mathbf{r},\mathbf{r},\omega)\cdot\hat{\mathbf{n}},
\end{equation}
where $\hat{\mathbf{n}}$ is the orientation of the dipole, $\mathbf{r}$ is its position, and $\omega$ is its frequency.

The dyadic Green function is computed using a quasi-electrostatic boundary element method (as in Ref.~\citenum{jin2017infrared}). In Figure~\ref{fig:3}, we plot the enhancement of the spontaneous emission rate $\Gamma/\Gamma_0$ for an external emitter polarized perpendicularly to the plane of the disk and placed \SI{5}{\nm} away from the center of the disk. Due to the orientation and position of the dipole , which maintains the axial symmetry of the disk, the emitter only couples to axially symmetric ($\ell = 0$)  modes with zero orbital angular momentum. We find that the rate of spontaneous emission of 2D optical phonons is approximately 8 orders of magnitude larger than the rate of spontaneous emission of photons in the far field at frequencies corresponding to resonant modes of the hBN disk. Such enhancement is much larger than the enhancement presented by an unstructured, infinite sheet at the same frequency, due to the concentration of electromagnetic local density of states around the resonances. Nevertheless, the average spontaneous emission enhancement, defined by the integral of the enhancement over frequencies, is comparable to that of the flat sheet, in keeping with sum rules for spontaneous emission enhancement~\cite{scheel2008macroscopic}.  We note that in this approach, the coupling of the dipole to phonon polaritons is manifested through the phonon contribution to the dielectric function of the disk. This should be equivalent to an approach that considers the coupling of a bound electron in an emitter to LO phonons in the disk through a 2D Fr\"ohlich coupling---\ie a coupling of the atomic electron to the electric potential resulting from the polarization associated with an LO phonon mode~\cite{sohier2016two}.

In Figure~\ref{fig:3}, we show that for an infrared emitter at a transition wavelength of \SI{7}{\micro\m} with a free-space radiative lifetime of \SI{1}{\micro\s}, \SI{5}{\nm} away from an hBN disk, the coupling rate to 2D optical phonons (about \SI{65}{\meV}) would be on the same scale as the optical phonon frequency itself (about \SI{180}{\meV}). This rate thus implies coupling between an emitter and the field in the regime of ultra-strong coupling. Moreover, the coupling rate for the \SI{20}{\nm} disk (purple), for an emitter with a far field decay rate of ${\gtrsim}\,\SI{3e14}{\per\s}$ would have a sufficient coupling strength to the distinct, radially-quantized resonances in purple for its linewidth to span multiple resonances and thus be in a multi-mode ultra-strong coupling regime. Thus, the extreme confinement of electromagnetic energy associated with LO phonons in two dimensions enables the possibility of realizing ultra-strong coupling of an atom or molecule with optical phonons in a polar material, allowing the potential realization of new coupled states of quantum emitters and phonons such as atom--phonon polariton bound states.

The ability to probe low-loss and highly confined electromagnetic modes associated with optical phonons in 2D polar materials provides a new platform for nanophotonics in the mid- and far-infrared spectral range. The identification of the phonon polariton of bulk and thin-film geometries with the 2D LO phonon made in this manuscript would extend the rich phenomenology of optical phonons to nanophotonic applications. This work also points the way to useful new approaches to study LO phonons, arising from the fact that 2D LO phonons, unlike their 3D counterparts, have their electromagnetic energy extend a considerable distance from the material boundary. Due to the strong electromagnetic interactions between emitters and 2D phonon polaritons shown here, it is now possible to design interesting new hybrid states of matter and phonons based on quantum electrodynamical strong coupling. The highly confined phonon polaritons in polar monolayers may also provide interesting new opportunities in near-field radiative heat transfer, in which it has been long known that thin-film surface phonon polaritons play a critical role. Additional opportunities come from considering the near- and far-field optical properties of periodically structured layers involving hBN and other materials such as graphene~\cite{papadakis2017ultralight}. An important avenue of future study would be the \emph{ab initio} calculation of lifetimes of 2D LO phonons associated with three-phonon processes~\cite{caldwell2015low,srivastava1990physics} and electron-phonon interactions~\cite{sundararaman2014theoretical, sohier2016two, ciccarino2018dynamics}. In further work, it would be of great interest to study the effects of isotopic purification and cryogenic temperatures on reducing the decay rate of these 2D LO phonons~\cite{giles2017ultra}.

\section{Acknowledgments}
The authors thank Ido Kaminer, Siyuan Dai, Samuel Moore, Jennifer Coulter, and Christopher Ciccarino for helpful discussions.
N.\,{}R. recognizes the support of the DOE Computational Science Graduate Fellowship (CSGF) fellowship no.\ DE-FG02-97ER25308. 
T.\,{}C. acknowledges support from the Danish Council for Independent Research (Grant No.\ DFF--6108-00667). 
This work was supported by the DOE Photonics at Thermodynamic Limits Energy Frontier Research Center under grant number DE-SC0019140.

\bibliographystyle{apsrev4-1}
\bibliography{references}

\end{document}